\documentclass[aps,prl,superscriptaddress,showpacs,twocolumn]{revtex4-1}

\usepackage{amsmath,amssymb,amstext}
\usepackage{graphicx,breakurl,color}
\usepackage[T1]{fontenc}
\usepackage[latin9]{inputenc}
\usepackage[normalem]{ulem}
\usepackage[unicode=true,pdfusetitle,pdfborder={0 0 1}]{hyperref}
\usepackage{upgreek}
\usepackage[english]{babel}

\hypersetup{colorlinks,
linkcolor=blue,          citecolor=blue,        filecolor=blue,      urlcolor=blue           }

\DeclareMathOperator{\tr}{tr}
\graphicspath{{../figures/}}

\begin{document}

\title{Quantum Computation under Micromotion in a Planar Ion Crystal}

\author{S.-T. Wang}
\email[Correspondence and requests for materials should be addressed to S.-T.W. ]{(wangst@umich.edu)}
\affiliation{Department of Physics, University of Michigan, Ann Arbor,
Michigan 48109, USA}
\affiliation{Center for Quantum Information, IIIS, Tsinghua University, Beijing 100084,
PR China}

\author{C. Shen}
\affiliation{Department of Physics, University of Michigan, Ann Arbor,
Michigan 48109, USA}
\affiliation{Department of Applied Physics, Yale University, New Haven,
Connecticut 06511, USA}

\author{L.-M. Duan}
\affiliation{Department of Physics, University of Michigan, Ann Arbor,
Michigan 48109, USA}
\affiliation{Center for Quantum Information, IIIS, Tsinghua University, Beijing 100084,
PR China}

\begin{abstract}
We propose a scheme to realize scalable quantum computation in a planar ion crystal confined by a Paul trap. We show that the inevitable in-plane micromotion affects the gate design via three separate effects: renormalization of the equilibrium positions, coupling to the transverse motional modes, and amplitude modulation in the addressing beam. We demonstrate that all of these effects can be taken into account and high-fidelity gates are possible in the presence of micromotion. This proposal opens the prospect to realize large-scale fault-tolerant quantum computation within a single Paul trap.
\end{abstract}

\maketitle





Scalable quantum computation constitutes one of the ultimate goals in modern
physics \cite{nielsen2010quantum, Ladd2010Quantum}. Towards that goal,
trapped atomic ions are hailed as one of the most promising platforms for
the eventual realization \cite{Blatt2008Entangled, Haffner2008Quantum}. The linear Paul trap with an one-dimensional (1D) ion crystal was among the first to perform quantum logic gates \cite{Cirac1995Quantum, Monroe1995Demonstration, Schmidt-Kaler2003Realization} and to generate entangled states \cite{Turchette1998Deterministic, Sackett2000Experimental, Roos2004Science}, but in terms of scalability, the 1D geometry limits the number of ions that can be successfully trapped \cite{Raizen1992Ionic, Schiffer1993Phase}. Another shortcoming of the 1D architecture is that the error threshold for fault-tolerant quantum computation with short-range gates is exceptionally low and very hard to be met experimentally \cite{Gottesman2000Fault, Svore2005Local, Szkopek2006Threshold}.

Generic ion traps, on the other hand, could confine up to millions
of ions with a 2D or 3D structure \cite{Itano1998Bragg, Drewsen1998Large, Mortensen2006Observation}. More crucially, large scale
fault-tolerant quantum computation can be performed with a high error
threshold, in the order of a percent level, with just nearest neighbor (NN)
quantum gates \cite{Raussendorf2007Fault, Raussendorf2007Topological, Fowler2009High,DiVincenzo2009Fault}. This
makes 2D or 3D ion crystals especially desirable for scalable quantum
computation. Various 2D architectures have been proposed, including
microtrap arrays \cite{Cirac2000Scalable}, Penning traps \cite%
{Porras2006Quantum, Zou2010Implementation, Itano1998Bragg,
Mitchell1998Direct}, and multizone trap arrays \cite%
{Kielpinski2002Architecture, Monroe2013Scaling}. However, the ion separation
distance in microtraps and penning traps is typically too large for fast
quantum gates since the effective ion-qubit interaction scales down rapidly
with the distance. In addition, fast rotation of the ion crystal in the
Penning trap makes the individual addressing of qubits very demanding. Distinct from these challenges, Paul traps provide
strong confinement; however, they are hampered by the micromotion problem: fast micromotion caused by the driving radio-frequency (rf) field cannot be laser cooled. It may thus create motion of large amplitudes well beyond the Lamb-Dicke regime \cite{berkeland1998minimization, Leibfried2003Quantum}, which becomes a serious impediment to high-fidelity quantum gates. 

In this paper, we propose a scheme for scalable quantum computation with a 2D ion crystal in a quadrupole Paul trap. We have shown recently that micromotion may not be an obstacle for design of high-fidelity gates for the two-ion case \cite{Shen2014High}. Here, we extend this idea and show that micromotion can be explicitly taken into account in the design of
quantum gates in a large ion crystal. This hence clears the critical hurdle
and put Paul traps as a viable architecture to realize scalable quantum
computation. In such a trap, DC and AC electrode voltages can be adjusted so
that a planar ion crystal is formed with a strong trapping potential in the
axial direction. In-plane micromotion is significant, but essentially no
transverse micromotion is excited due to negligible displacement from the
axial plane. We perform gates mediated by transverse motional modes and show that the in-plane micromotion influences the gate design through three
separate ways: (1) It renormalizes the average positions of each ion compared
to the static pseudopotential equilibrium positions. (2) It couples to and
modifies the transverse motional modes. (3) It causes amplitude modulation in
the addressing beam. In contrast to thermal motion, the fluctuation induced
by micromotion is coherent and can be taken into account explicitly. Several other works also studied the effect of micromotion on equilibrium ion positions and motional modes \cite{Landa2012Modes, Kaufmann2012Precise, Landa2014Entanglement}, or used transverse modes in an oblate Paul trap to minimize the micromotion effect \cite{Yoshimura2014Creation}.  
Here, by using multiple-segment laser pulses \cite{Zhu2006Arbitrary, Zhu2006Trapped, Choi2014Optimal}, we demonstrate that high-fidelity quantum gates can be achieved even in the presence of
significant micromotion and even when many motional modes are excited. Our work therefore shows the feasibility of quadrupole Paul traps in performing large scale quantum computation, which may drive substantial experimental progress.

A generic quadrupole Paul trap can be formed by electrodes with a hyperbolic
cross-section. The trap potential can be written as $\Phi (x,\,y,\,z)=\Phi _{%
\text{DC}}(x,\,y,\,z)+\Phi _{\text{AC}}(x,\,y,\,z)$, where
\begin{align}
\Phi _{\text{DC}}(x,\,y,\,z)& =\frac{U_{0}}{d_{0}^{2}}\left[ (1+\gamma
)x^{2}+(1-\gamma )y^{2}-2z^{2}\right] , \\
\Phi _{\text{AC}}(x,\,y,\,z)& =\frac{V_{0}\cos (\Omega _{T}t)}{d_{0}^{2}}%
\left( x^{2}+y^{2}-2z^{2}\right) .
\end{align}%
It contains both a DC and an AC part, with $U_{0}$ being the DC voltage, and
$V_{0}$ being the AC voltage forming an electric field oscillating at the
radiofrequency $\Omega _{T}$. The parameter $d_{0}$ characterizes the size
of the trap and $\gamma $ controls the anisotropy of the potential in the $x$%
-$y$ plane. We choose $\gamma $ to deviate slightly from zero, so that the
crystal cannot rotate freely in the plane, i.e.\ to remove the gapless
rotational mode. The AC part, on the contrary, is chosen to be isotropic in
the $x$-$y$ plane. We let $U_{0}<0$ such that the trapping is enhanced along
the $z$ direction in order to form a 2D crystal in the $x$-$y$ plane.
Disregarding the Coulomb potential first, the equations of motion of ions in
such a trap can be written in the standard form of Mathieu equations along
each direction:
\begin{equation}
\frac{d^{2}r_{\nu }}{d\xi ^{2}}+\left[ a_{\nu }-2q_{\nu }\cos (2\xi )\right]
r_{\nu }=0,
\end{equation}%
where $\nu \in \{x,y,z\}$, and the dimensionless parameters are $\xi =\Omega
_{T}t/2$, $a_{x}=8(1+\gamma )eU_{0}/md_{0}^{2}\Omega _{T}^{2}$, $%
a_{y}=8(1-\gamma )eU_{0}/md_{0}^{2}\Omega _{T}^{2}$, $%
a_{z}=-16eU_{0}/md_{0}^{2}\Omega _{T}^{2}$, $%
q_{x}=q_{y}=q=-4eV_{0}/md_{0}^{2}\Omega _{T}^{2}$, $q_{z}=-2q$. Neglecting
micromotion, one could approximate the potential as a time-independent
harmonic pseudopotential with secular trapping frequencies $\omega _{\nu
}=\beta _{\nu }\Omega _{T}/2$, with $\beta _{\nu }\approx \sqrt{a_{\nu} +q_{\nu}^{2}/2}$ being the
characteristic exponents of the Mathieu equations \cite{mclachlan1951theory,
king1999quantum}. \\

\noindent {\large \textbf{Results}}\\
\textbf{Dynamic ion positions.} 
Adding Coulomb interactions back, the static equilibrium positions can be found by minimizing the total pseudopotential \cite{james1998quantum, Zou2010Implementation}, or use molecular dynamics simulation with added dissipation, which imitates the cooling process in experiment \cite{Zhang2007Molecular, Schiffer2000Temperature}. In our
numerical simulation, we start with $N=127$ ions forming equilateral
triangles in a 2D hexagonal structure. We then solve
the equations of motion with a small frictional force to find the
equilibrium positions $\vec{r}\,^{(0)} =\vec{r}(t\to\infty) =
(x_{1}^{(0)},y_{1}^{(0)}, \cdots, x_{N}^{(0)},y_{N}^{(0)})$, which is the
starting point for the expansion of the Coulomb potential. Micromotion is
subsequently incorporated by solving the decoupled driven Mathieu equations (see supplementary materials). The average ion positions $\vec{r}\,^{(0)}$ are found
self-consistently, which differ slightly from the pseudopotential
equilibrium positions (an average of $0.03 \, \mu$m shift). Dynamic ion
positions $\vec{r}(t)$ can be expanded successively as
\begin{equation}
\vec{r}(t) = \vec{r}\,^{(0)} + \vec{r}\,^{(1)} \cos(\Omega_{T} t) + \vec{r}%
\,^{(2)} \cos(2\Omega_{T} t) + \cdots.  \label{Eq:Positions}
\end{equation}
Numerically, we found that $\vec{r}\,^{(1)} \approx -\frac{q}{2} \vec{r}%
\,^{(0)}$ and $\vec{r}\,^{(2)} \approx \frac{q^{2}}{32} \vec{r}\,^{(0)}$,
where the expression for $\vec{r}\,^{(1)}$ is consistent with previous
results \cite{Shen2014High, Landa2012Modes, Zhang2007Molecular}. Micromotion
thus only results in breathing oscillations about the average positions.

\begin{figure}[t]
\hspace{-.2cm} \includegraphics[trim=.1cm 0cm 2.3cm 1.5cm,
clip,width=0.48\textwidth]{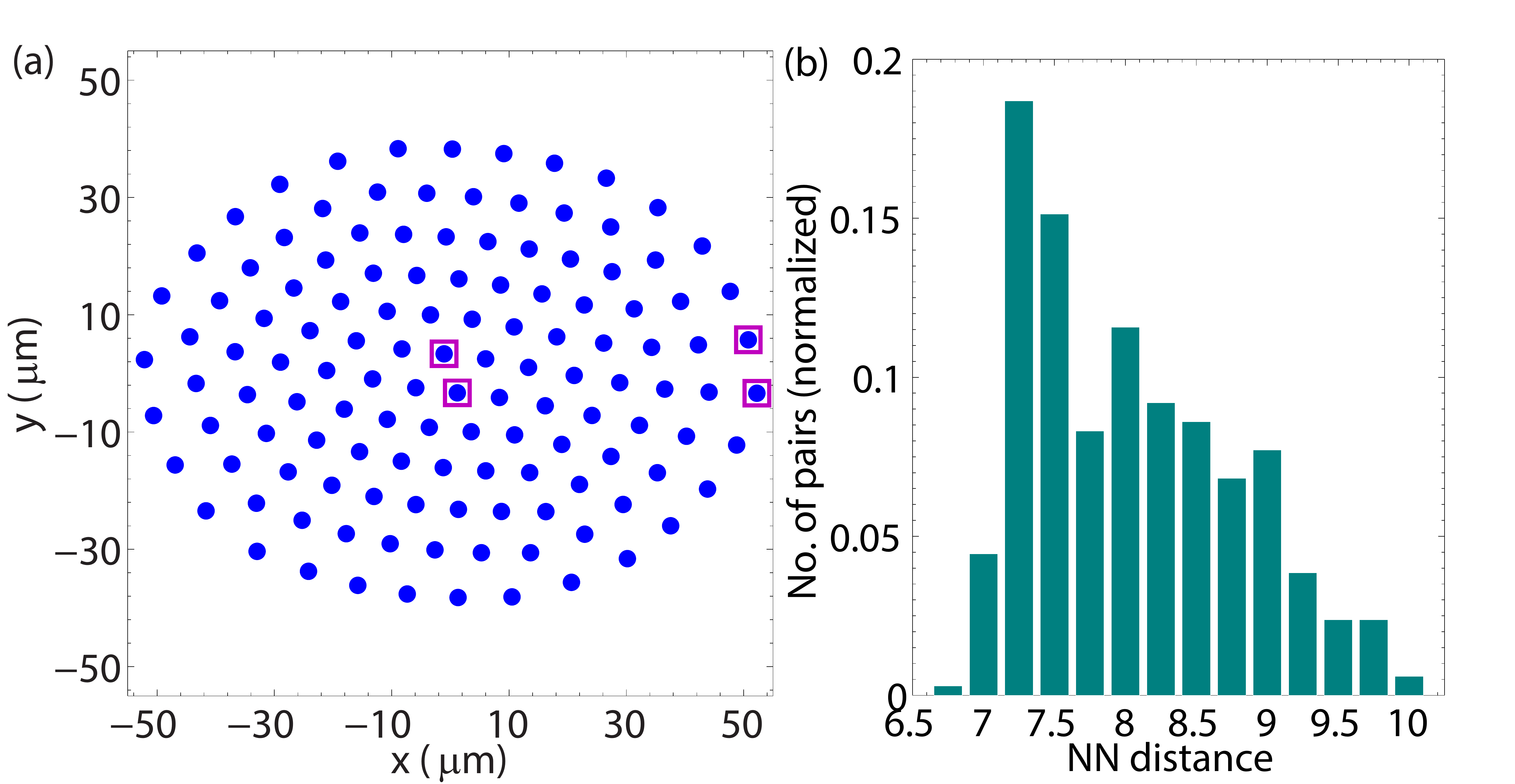}
\caption{ \textbf{Crystal structure and distance distribution.} (a) Average positions $\vec{r}\,^{(0)}$ of 127 ions
in a planar crystal. Breathing oscillations about these average positions
occur due to micromotion. Two pairs of ions (enclosed in squares), one pair
in the center and one near the edge, are used for the demonstration of a
quantum gate later. (b) The distribution of nearest neighbor (NN) distance.
The minimum, maximum, and average NN distances are $6.9\,\protect\mu$m, $10\,%
\protect\mu$m and $8.0\,\protect\mu$m respectively. Parameters used are: the number of ions $N=127$; DC and AC potential $U_{0}=-1.1\,$V, $V_{0}=90\,$V; AC rf frequency $\Omega_{T}/2\protect\pi=50\,$MHz; the
characteristic electrode size $d_{0}=200\,\protect\mu$m; ion mass $m=171u$ ($%
u$ is the atomic mass unit) corresponds to $^{171}$Yb$^{+}$ ion; the
anisotropy parameter $\protect\gamma=0.01$; corresponding Mathieu parameters
are $a_{x}\approx-1.27 \times 10^{-3}, a_{y}\approx-1.25 \times 10^{-3},
a_{z}\approx 2.52 \times 10^{-3}, q\approx-0.051$, with respective secular
trap frequencies $\protect\omega_{x}/2\protect\pi\approx0.18\,$MHz, $\protect%
\omega_{y}/2\protect\pi\approx0.22\,$MHz, $\protect\omega_{z}/2\protect\pi%
\approx2.21\,$MHz; $\protect\omega_{z}/\protect\omega_{x,y}>10$ ensures a planar crystal is formed.}
\label{Fig:IonPosition}
\end{figure}

Fig.\ \ref{Fig:IonPosition}(a) shows the average ion positions $\vec{r}%
\,^{(0)}$ in the planar crystal. The distribution of NN distance is plotted
in figure \ref{Fig:IonPosition}(b). We choose the voltages $U_{0}$ and $V_{0}
$ such that the ion distance is kept between $6.5\,\mu $m and $10\,\mu $m.
This ensures that crosstalk errors due to the Gaussian profile of the
addressing beam are negligible, at the same time maintaining strong
interaction between the ions. As micromotion yields breathing oscillations,
the further away the ion is from the trap center, the larger the amplitude
of micromotion becomes. With the furthest ion around $52\,\mu $m from the
trap center, the amplitude of micromotion is $-q/2\times 52\approx 1.4\,\mu $m, which is well below the separation distance between the ions but larger than the optical wavelength (see supplementary materials for the distribution of the amplitude of micromotion). \\

\noindent \textbf{Normal modes in the transverse direction.}
With the knowledge of ion motion in the $x$-$y$ plane, we proceed to find the normal modes and quantize the motion along the transverse ($z$) direction. As ions are confined in the plane, micromotion along the transverse direction is negligible. The harmonic pseudopotential approximation is therefore legitimate. Expanding the Coulomb potential to second order, we have $\frac{\partial ^{2}}{\partial
z_{i}\partial z_{j}}\Big(\frac{1}{\tilde{r}_{ij}}\Big)\Bigr\rvert_{\vec{r}%
(t)}=\frac{1}{r_{ij}^{3}}$, where $\tilde{r}_{ij}=\sqrt{%
(x_{i}-x_{j})^{2}+(y_{i}-y_{j})^{2}+(z_{i}-z_{j})^{2}}$ is the 3D distance
and $r_{ij}=\sqrt{(x_{i}-x_{j})^{2}+(y_{i}-y_{j})^{2}}$ is the planar
distance between ions $i$ and $j$. To the second order, transverse and
in-plane normal modes are decoupled. Note that coupling between
the in-plane micromotion and the transverse normal modes has been taken
into account in this expansion as the Coulomb potential is expanded around
the dynamic ion positions $\vec{r}(t)$. With significant in-plane
micromotion, distances between ions are time-dependent, which in turn
affects the transverse modes. We can expand the quadratic coefficients in
series:
\begin{equation}
\dfrac{1}{r_{ij}^{3}}\approx \Big<\dfrac{1}{r_{ij}^{3}}\Big>+M_{ij}\cos
(\Omega _{T}t)+\cdots .
\end{equation}%
The time-averaged coefficients $\left\langle 1/r_{ij}^{3}\right\rangle $ can
be used to compute the transverse normal modes. The next order containing $%
\cos (\Omega _{T}t)$ terms can be considered as a time-dependent
perturbation to the Hamiltonian. It contributes on the order of $O\left(
q\omega _{k}^{2}/\Omega _{T}^{2}\right) \sim O(qq_{z}^{2})$ in the rotating
wave approximation, where $\omega _{k}$ is the transverse mode frequency. The term $\left\langle 1/r_{ij}^{3}\right\rangle
\approx \big(1/r_{ij}^{(0)}\big)^{3}(1-3q^{2}/4)+O(q^{3})$, where $%
r_{ij}^{(0)}$ is the ion distance computed with $\vec{r}\,^{(0)}$ without
considering micromotion (see supplementary materials). Here, the micromotion effect
is an overall renormalization in the term $1/r_{ij}^{3}$, so it does not
modify the normal mode structure. Instead, it slightly shifts down the
transverse mode frequencies (in the order of $O(q^{2})$). Numerically, we found an average reduction of around $0.4\,$kHz in each transverse mode frequency with our chosen parameters. Although mode structure is not altered by this overall renormalization, the discrepancy in equilibrium positions compared to the pseudopotential approximation will modify both the normal mode structure and mode frequencies.

\begin{figure}[t]
\includegraphics[trim=0cm 0cm 0cm 0cm, clip,width=0.47\textwidth]{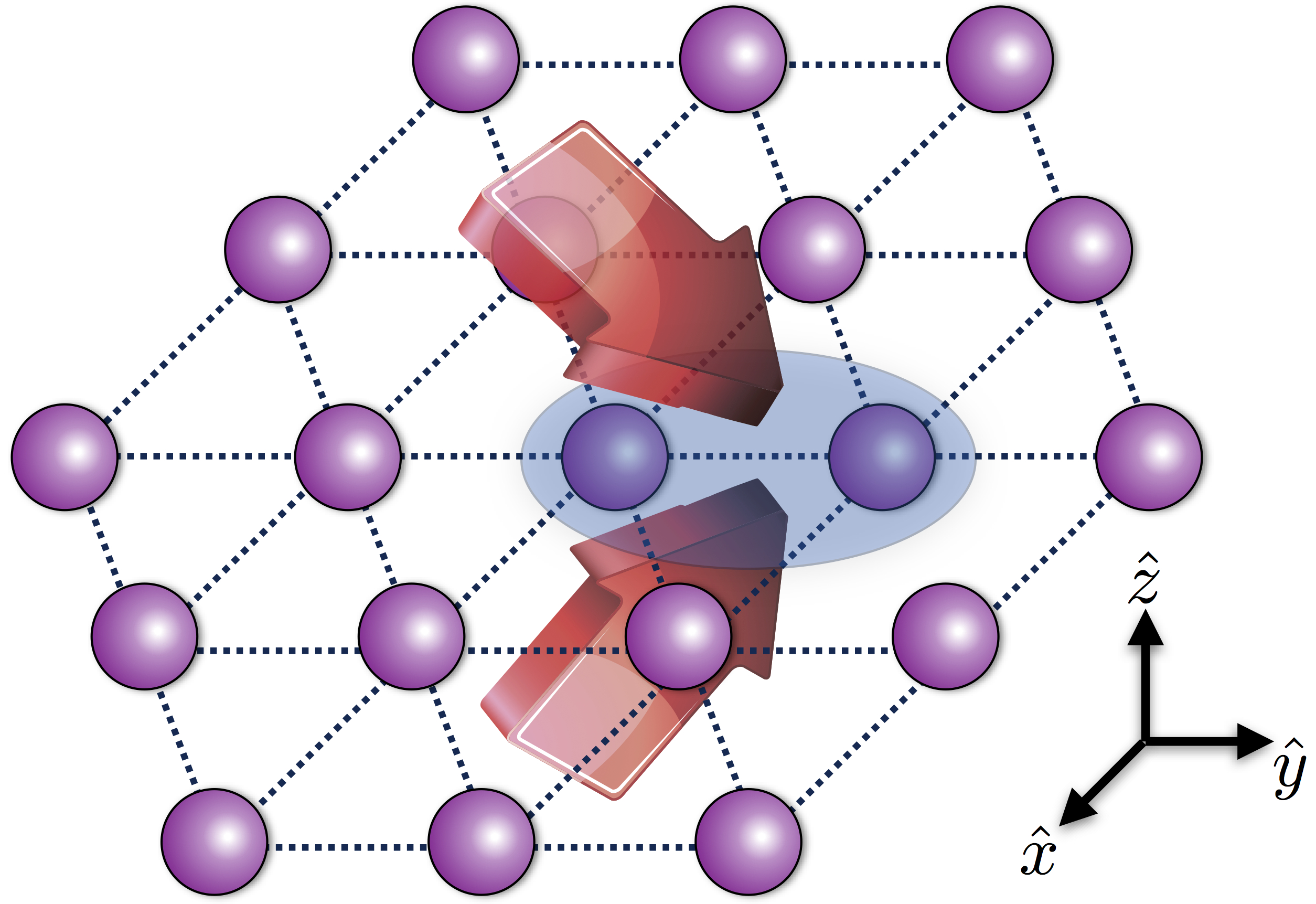}
\caption{ \textbf{Nearest neighbor quantum gate in a 2D planar crystal.} Two laser beams with a wave vector difference $\Delta k$ aligned in the $z$ direction exert a spin-dependent force on the neighboring ions. Parameters used are: The wave vector difference of addressing beams $\Delta k=8 \, \protect\mu\text{m}^{-1}$; Laser beams are assumed to take a Gaussian profile with a beam waist $w=3 \, \mu$m centered at the average positions of the respective ion; The Lamb-Dicke parameter $\protect\eta_{z} = \Delta k \protect\sqrt{
\hbar/2m\protect\omega_{z} } \approx 0.029$. Other parameters are the same as in Fig.\ \ref{Fig:IonPosition}.}
\label{Fig:Schematics}
\end{figure}

\mbox{}\newline
\textbf{High-fidelity quantum gates.}
After obtaining the correct transverse normal modes, we now show how to design high-fidelity quantum gates with in-plane micromotion. Since NN gates are sufficient for fault-tolerant quantum computation in a planar crystal, we show as a demonstration that high-fidelity entangling gates can be achieved with a pair of NN ions in the trap center and near the trap edge. One may perform the gate along the transverse direction by shining two laser beams on the two NN ions with wave vector difference $\Delta k\hat{z}$ and frequency difference $\mu $ (see Fig.\ \ref{Fig:Schematics}) \cite{Leibfried2003Experimental, Choi2014Optimal}. The laser-ion interaction Hamiltonian is \cite{Zhu2006Trapped} $H=\sum_{j=1}^{2}\hbar \Omega _{j}\cos (\Delta k \cdot \delta z_{j}+\mu t)\sigma _{j}^{z}$ , where $\Omega _{j}$ is the (real) Raman Rabi frequency for the $j$th ion, $\sigma _{j}^{z}$ is the Pauli-$Z$ matrix acting on the pseudospin space of internal atomic states of the ion $j$, and $\delta z_{j}$ is the ion displacement from the equilibrium position. Quantize the ion motion, $\delta z_{j}=\sum_{k}\sqrt{\hbar /2m\omega _{k}}b_{j}^{k}(a_{k}+a_{k}^{\dag })$, with $b_{j}^{k}$ ($\omega _{k}$) being the mode vector (frequency) for mode $k$ and $a_{k}^{\dagger }$ creates the $k$-th phonon mode. Expanding the cosine term and ignoring the single-bit operation, the Hamiltonian can be written in the interaction picture as 
\begin{equation}
H_{\text{I}}= - \sum_{j=1}^{2} \sum_{k} \chi_{j} (t) g_{j}^{k} \big(a^{\dag}_{k} e^{i \omega_{k} t}+ a_{k} e^{-i \omega_{k} t} \big) \sigma_{j}^{z},
\end{equation}
where $\chi_{j} (t) = \hbar \Omega_{j} \sin (\mu t)$, $g_{j}^{k} = \eta_{k} b_{j}^{k}$, and the Lamb-Dicke parameter $\eta_{k}= \Delta k \sqrt{ \hbar/2m\omega_{k} } \ll 1$. The evolution operator corresponding to the Hamiltonian $H_{\text{I}}$ can be written as \cite{Zhu2006Trapped,Kim2009Entanglement, Choi2014Optimal}
\begin{equation}
U(\tau) = \exp\Big( i \sum_{j} \phi_{j} (\tau) \sigma_{j}^{z} + i \sum_{j<n} \phi_{jn} (\tau) \sigma_{j}^{z} \sigma_{n}^{z} \Big), \label{Eq:Evolution}
\end{equation}
where the qubit-motion coupling term
$\phi_{j} (\tau) =-i \sum_{k} \alpha_{j}^{k} (\tau) a_{k}^{\dag}- \alpha_{j}^{k*}(\tau) a_{k}$
with 
$\alpha_{j}^{k}(\tau) = \frac{i}{\hbar} g_{j}^{k} \int_{0}^{\tau} \chi_{j}(t) e^{i \omega_{k}t} dt$ and the two-qubit conditional phase 
$\phi_{jn}(\tau) = \frac{2}{\hbar^{2}} \sum_{k} g_{j}^{k}g_{n}^{k}  \int_{0}^{\tau} \int_{0}^{t_{2}} \chi_{j} (t_{2}) \times \chi_{n}(t_{1}) \sin(\omega_{k}(t_{2}-t_{1}) ) dt_{1} dt_{2}$. To realize a conditional phase flip (CPF) gate between ions $j$ and $n$, we require $\alpha _{j}^{k}\approx 0$ so that the spin and phonons are almost disentangled at the end of the gate, and also $\phi _{jn}(\tau )=\pi /4$. It is worthwhile to note that in deriving Eq.\ \eqref{Eq:Evolution}, we dropped single-qubit operations as we are interested in the CPF gate. These fixed single-qubit operations can be explicitly compensated in experiment by subsequent rotations of single spins. (see supplementary materials for more detailed derivation and analysis).

\begin{figure*}[t]
\includegraphics[trim=0cm 0cm 0cm 0cm, clip,width=\textwidth]{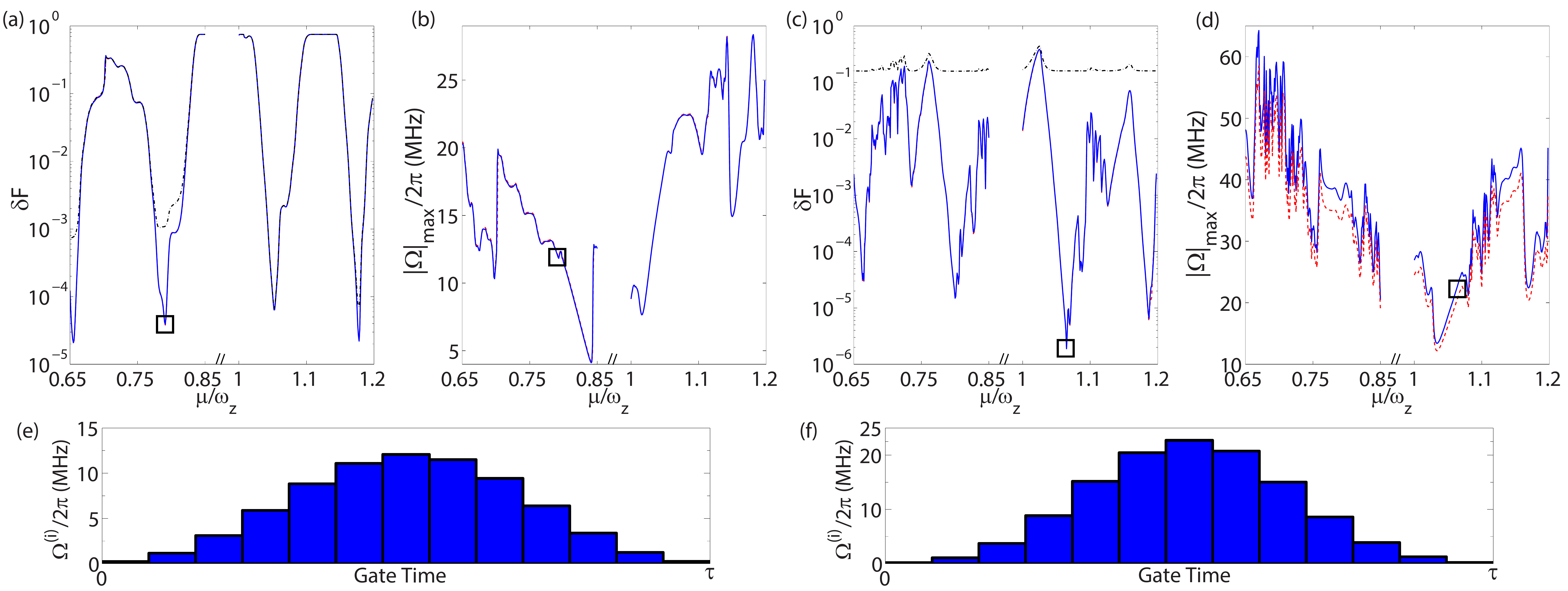} 
\caption{ \textbf{Gate infidelity and pulse shaping.} (a), (b), and (e) are respectively the gate infidelity, the maximum Rabi frequency, and the thirteen-segment pulse pattern corresponding to the results marked by squares, for the center pair as labeled in Fig.\ \protect\ref{Fig:IonPosition}. (c), (d), and (f) are the corresponding plots for the edge pair. The blue solid lines and the pulse
sequences indicate the optimal results with micromotion considered. The red
dashed lines are results for a genuine static harmonic trap without
micromotion. Black dash-dot lines in (a) and (c) are obtained by applying
the optimal solution for a static trap to the case with micromotion. All
transverse modes are distributed between $0.85\protect\omega_{z}$ and $\protect%
\omega_{z}$. We optimize the gate near either end of the spectrum. The
optimal results marked by the squares are $\protect\delta F= 4\times 10^{-5}$
and $|\Omega|_{\text{max}}/2\protect\pi=12 \,$MHz ($\protect\delta F=
4\times 10^{-6}$ and $|\Omega|_{\text{max}}/2\protect\pi=22 \,$MHz) for the
center (edge) pair. Parameters used are: total gate time $\protect\tau =
50\times 2\protect\pi/\protect\omega_{z} \approx 23 \, \protect\mu$s; $m=13$ segments are used; Doppler temperature $%
k_{B}T_{D}/\hbar \approx 2\protect\pi \times 10\,$MHz is assumed for all
phonon modes. Other parameters are the same as in Fig.\ \protect\ref%
{Fig:IonPosition} and Fig.\ \ref{Fig:Schematics}.}
\label{Fig:Gate}
\end{figure*}

As the number of ions increases, transverse phonon modes become very close
to each other in frequencies. During typical gate time, many motional modes
will be excited. We use multiple-segment pulses to achieve a high-fidelity
gate \cite{Zhu2006Arbitrary,Zhu2006Trapped}. The total gate time is divided
into $m$ equal-time segments, and the Rabi frequency takes the form $\Omega
_{j}(t)=\Omega _{j}^{(i)}\Omega _{j}^{\text{G}}(t)$, with $\Omega _{j}^{(i)}$
being the controllable and constant amplitude for the $i$th segment ($%
(i-1)\tau /m\leq t<i\tau /m$). Due to the in-plane micromotion, the laser
profile $\Omega _{j}^{\text{G}}(t)$ seen by the ion is time-dependent. In
our calculation, we assume the Raman beam to take a Gaussian form, with $%
\Omega _{j}^{\text{G}}(t)=\exp \left\{ -\left[ \big(x_{j}(t)-x_{j}^{(0)}\big)%
^{2}+\big(y_{j}(t)-y_{j}^{(0)}\big)^{2}\right] /w^{2}\right\} $, where $w$
is the beam waist and $\big(x_{j}^{(0)},y_{j}^{(0)}\big)$ are the average
positions for the $j$th ion. Any other beam profile can be similarly
incorporated.

To gauge the quality of the gate, we use a typical initial state for the ion
spin $|\Phi _{0}\rangle =\left( |0\rangle +|1\rangle \right) \otimes \left(
|0\rangle +|1\rangle \right) /2$ and the thermal state $\rho _{m}$ for the
phonon modes at the Doppler temperature. The fidelity is defined as $F=\tr%
_{m}\left[ \rho _{m}\big|\langle \Psi _{0}|U_{\text{CPF}}^{\dagger }U(\tau
)|\Psi _{0}\rangle \big|^{2}\right] $ tracing over the phonon modes, with
the evolution operator $U(\tau )$ and the perfect CPF gate $U_{\text{CPF}%
}\equiv e^{i\pi \sigma _{1}^{z}\sigma _{2}^{z}}$. For simplicity, we take $%
\Omega _{j}^{(i)}=\Omega _{n}^{(i)}=\Omega ^{(i)}$ for the ions $j$ and $n$.
For any given detuning $\mu $ and gate time $\tau $, we optimize the control parameters $\Omega ^{(i)}$ to get the maximum fidelity $F$. Fig.\ \ref{Fig:Gate} shows the gate infidelity $\delta F=1-F$ and the maximum Rabi
frequency $|\Omega |_{\max }=\max_{i}\Omega ^{(i)}$ for the center pair [(a)
and (b)] and the edge pair [(c) and (d)] with $13$ segments and a relatively
fast gate $\tau \approx 23\,\mu $s. Detuning $\mu$ can be used as an adjusting parameter in experiment to find the optimal results. All transverse phonon modes are distributed between $0.85 \omega_{z}$ and $\omega_{z}$. We optimize the gate near either end of the spectrum since optimal results typically occur there. Blue solid lines indicate the optimal
results with micromotion and red dashed lines show the results for a genuine
static harmonic trap, which are almost identical in (a), (b) and (c). It
implies that micromotion can almost be completely compensated, but with a
stronger laser power for the edge pair. If we apply the optimal result for
the static trap to the realistic case with micromotion, the fidelity will be
lower as indicated by the black dash-dot lines. This is especially so for
the edge pair, where the fidelity is lower than $85\%$ at any detuning. It
is therefore critical to properly include the effect of micromotion. With
corrected pulse sequences, a fidelity $F>99.99\%$ can be attained with $%
|\Omega |_{\max }/2\pi \approx 12\,$MHz ($|\Omega |_{\max }/2\pi \approx 22\,
$MHz) for the center (edge) ions. The Rabi frequencies can be further
reduced by a slower gate and/or more pulse segments. \\
\newline
\textbf{Noise estimation.} 
Micromotion of any amplitude does not induce errors to the gates as it has been completely compensated in our gate design. We now estimate various other sources of noise for gate implementation. In considering the effect of in-plane micromotion to the transverse modes, we are accurate to the order of $q^{2}$, so an error of $q^{3}\approx 10^{-4}$ is incurred. The actual error is smaller since the Coulomb potential is an order of magnitude smaller than the trapping potential along the transverse direction. The cross-talk error probability
due to beam spillover is $P_{c}=e^{-2(d/w)^{2}}<2\times 10^{-5}$, with the
ion distance $d\gtrsim 7\,\mu $m and the beam waist $w=3\,\mu $m. At the
Doppler temperature $k_{B}T_{D}/\hbar \approx 2\pi \times 10\,$MHz, thermal
spread in positions may degrade the gate fidelity. Similar to micromotion,
thermal motion causes the effective Rabi frequency to fluctuate. With $%
\omega _{x,y}/2\pi \approx 0.2\,$MHz, there is a mean phonon number $\bar{n}%
_{0}\approx 50$ in the $x$-$y$ plane. It gives rise to thermal motion with average fluctuation in positions, $\delta r\approx 0.23\,\mu $m, which can be estimated as in Ref.\ \onlinecite{Lin2009Large}. The resultant gate infidelity is $\delta F_{1} \approx (\pi^{2}/4)(\delta r/w)^{4} \approx 10^{-4}$. Lastly, we estimate the infidelity caused by higher-order expansion in the Lamb-Dicke parameter. The infidelity is $\delta F_{2}\approx \pi
^{2}\eta _{z}^{4}(\bar{n}_{z}^{2}+\bar{n}_{z}+1/8)\approx 2\times 10^{-4}$,
where $\bar{n}_{z}\approx 5$ is the mean phonon number in the transverse
direction \cite{Zhu2006Trapped}. Other than the effects considered above,
micromotion may also lead to rf heating when it is coupled to thermal
motion. However, simulation has shown that at low temperature $T<10\,$mK and
small $q$ parameters, rf heating is negligible \cite%
{Zhang2007Molecular,Ryjkov2005Simulations}. Heating effect due to rf phase shift and voltage fluctuation should also be negligible when they are well-controlled \cite{Zhang2007Molecular}. \\

\noindent {\large \textbf{Discussion}}\\
It is worthwhile to point out that although we have demonstrated the feasibility of our gate design via a single case with $N=127$ ions, the proposed scheme scales for larger crystals. The intuition is that through optimization of the segmented pulses, all phonon modes are nearly disentangled from the quantum qubits at the end of the gate. However, as the number of ions further increases, one would presumably need more and more precise control for all the experimental parameters ($<1\%$ fluctuation in voltage for example). rf heating may also destabilize a much larger crystal \cite{Buluta2008Investigation}, and more careful studies are necessary for larger crystals.   

One may also notice that in Ref.\ \onlinecite{Shen2014High}, we considered gates mediated by the longitudinal phonon modes, so the effect of micromotion is a phase modulation. Here, we utilize transverse modes so the amplitude of the laser beam is modulated. There are a few advantages in using the transverse modes: first, it is experimentally easier to access the transverse phonon modes in a planar ion crystal; second, in a planar crystal, the transverse direction is tightly trapped, so micromotion along that direction can be neglected; third, the transverse phonon modes do not couple to the in-plane modes and the in-plane micromotion affects the transverse modes via the time-dependence of the equilibrium positions, the effect of which is again suppressed due to tight trapping in the transverse direction.

In summary, we have demonstrated that a planar ion crystal in a quadrupole Paul trap is a promising platform to realize scalable quantum computation when micromotion is taken into account explicitly. We show that the in-plane micromotion comes into play through three separate effects, and each of them can be resolved. This paves a new pathway for large-scale trapped-ion quantum computation.


\begin{thebibliography}{10}
\expandafter\ifx\csname url\endcsname\relax
  \def\url#1{\texttt{#1}}\fi
\expandafter\ifx\csname urlprefix\endcsname\relax\def\urlprefix{URL }\fi
\providecommand{\bibinfo}[2]{#2}
\providecommand{\eprint}[2][]{\url{#2}}

\bibitem{nielsen2010quantum}
\bibinfo{author}{Nielsen, M.~A.} \& \bibinfo{author}{Chuang, I.~L.}
\newblock \emph{\bibinfo{title}{Quantum computation and quantum information}}
\newblock  (\bibinfo{publisher}{Cambridge university press},
  \bibinfo{year}{2010}).

\bibitem{Ladd2010Quantum}
\bibinfo{author}{Ladd, T.~D.} \emph{et~al.}
\newblock \bibinfo{title}{Quantum computers}.
\newblock \emph{\bibinfo{journal}{Nature}} \textbf{\bibinfo{volume}{464}},
  \bibinfo{pages}{45--53}
\newblock  (\bibinfo{year}{2010}).

\bibitem{Blatt2008Entangled}
\bibinfo{author}{Blatt, R.} \& \bibinfo{author}{Wineland, D.}
\newblock \bibinfo{title}{Entangled states of trapped atomic ions}.
\newblock \emph{\bibinfo{journal}{Nature}} \textbf{\bibinfo{volume}{453}},
  \bibinfo{pages}{1008--1015}
\newblock  (\bibinfo{year}{2008}).

\bibitem{Haffner2008Quantum}
\bibinfo{author}{Haffner, H.}, \bibinfo{author}{Roos, C.~F.} \&
  \bibinfo{author}{Blatt, R.}
\newblock \bibinfo{title}{Quantum computing with trapped ions}.
\newblock \emph{\bibinfo{journal}{Phys. Rep.}} \textbf{\bibinfo{volume}{469}},
  \bibinfo{pages}{155--203}
\newblock  (\bibinfo{year}{2008}).

\bibitem{Cirac1995Quantum}
\bibinfo{author}{Cirac, J.~I.} \& \bibinfo{author}{Zoller, P.}
\newblock \bibinfo{title}{Quantum computations with cold trapped ions}.
\newblock \emph{\bibinfo{journal}{Phys. Rev. Lett.}}
  \textbf{\bibinfo{volume}{74}}, \bibinfo{pages}{4091--4094}
\newblock  (\bibinfo{year}{1995}).

\bibitem{Monroe1995Demonstration}
\bibinfo{author}{Monroe, C.}, \bibinfo{author}{Meekhof, D.~M.},
  \bibinfo{author}{King, B.~E.}, \bibinfo{author}{Itano, W.~M.} \&
  \bibinfo{author}{Wineland, D.~J.}
\newblock \bibinfo{title}{Demonstration of a fundamental quantum logic gate}.
\newblock \emph{\bibinfo{journal}{Phys. Rev. Lett.}}
  \textbf{\bibinfo{volume}{75}}, \bibinfo{pages}{4714--4717}
\newblock  (\bibinfo{year}{1995}).

\bibitem{Schmidt-Kaler2003Realization}
\bibinfo{author}{Schmidt-Kaler, F.} \emph{et~al.}
\newblock \bibinfo{title}{Realization of the cirac-zoller controlled-not
  quantum gate}.
\newblock \emph{\bibinfo{journal}{Nature}} \textbf{\bibinfo{volume}{422}},
  \bibinfo{pages}{408--411}
\newblock  (\bibinfo{year}{2003}).

\bibitem{Turchette1998Deterministic}
\bibinfo{author}{Turchette, Q.~A.} \emph{et~al.}
\newblock \bibinfo{title}{Deterministic entanglement of two trapped ions}.
\newblock \emph{\bibinfo{journal}{Phys. Rev. Lett.}}
  \textbf{\bibinfo{volume}{81}}, \bibinfo{pages}{3631--3634}
\newblock  (\bibinfo{year}{1998}).

\bibitem{Sackett2000Experimental}
\bibinfo{author}{Sackett, C.~A.} \emph{et~al.}
\newblock \bibinfo{title}{Experimental entanglement of four particles}.
\newblock \emph{\bibinfo{journal}{Nature}} \textbf{\bibinfo{volume}{404}},
  \bibinfo{pages}{256--259}
\newblock  (\bibinfo{year}{2000}).

\bibitem{Roos2004Science}
\bibinfo{author}{Roos, C.~F.} \emph{et~al.}
\newblock \bibinfo{title}{Control and measurement of three-qubit entangled
  states}.
\newblock \emph{\bibinfo{journal}{Science}} \textbf{\bibinfo{volume}{304}},
  \bibinfo{pages}{1478--1480}
\newblock  (\bibinfo{year}{2004}).

\bibitem{Raizen1992Ionic}
\bibinfo{author}{Raizen, M.~G.}, \bibinfo{author}{Gilligan, J.~M.},
  \bibinfo{author}{Bergquist, J.~C.}, \bibinfo{author}{Itano, W.~M.} \&
  \bibinfo{author}{Wineland, D.~J.}
\newblock \bibinfo{title}{Ionic crystals in a linear paul trap}.
\newblock \emph{\bibinfo{journal}{Phys. Rev. A}} \textbf{\bibinfo{volume}{45}},
  \bibinfo{pages}{6493--6501}
\newblock  (\bibinfo{year}{1992}).

\bibitem{Schiffer1993Phase}
\bibinfo{author}{Schiffer, J.~P.}
\newblock \bibinfo{title}{Phase transitions in anisotropically confined ionic
  crystals}.
\newblock \emph{\bibinfo{journal}{Phys. Rev. Lett.}}
  \textbf{\bibinfo{volume}{70}}, \bibinfo{pages}{818--821}
\newblock  (\bibinfo{year}{1993}).

\bibitem{Gottesman2000Fault}
\bibinfo{author}{Gottesman, D.}
\newblock \bibinfo{title}{Fault-tolerant quantum computation with local gates}.
\newblock \emph{\bibinfo{journal}{J. Mod. Opt.}} \textbf{\bibinfo{volume}{47}},
  \bibinfo{pages}{333--345}
\newblock  (\bibinfo{year}{2000}).

\bibitem{Svore2005Local}
\bibinfo{author}{Svore, K.~M.}, \bibinfo{author}{Terhal, B.~M.} \&
  \bibinfo{author}{DiVincenzo, D.~P.}
\newblock \bibinfo{title}{Local fault-tolerant quantum computation}.
\newblock \emph{\bibinfo{journal}{Phys. Rev. A}} \textbf{\bibinfo{volume}{72}},
  \bibinfo{pages}{022317}
\newblock  (\bibinfo{year}{2005}).

\bibitem{Szkopek2006Threshold}
\bibinfo{author}{Szkopek, T.} \emph{et~al.}
\newblock \bibinfo{title}{Threshold error penalty for fault-tolerant quantum
  computation with nearest neighbor communication}.
\newblock \emph{\bibinfo{journal}{IEEE Trans. Nanotechnol.}}
  \textbf{\bibinfo{volume}{5}}, \bibinfo{pages}{42--49}
\newblock  (\bibinfo{year}{2006}).

\bibitem{Itano1998Bragg}
\bibinfo{author}{Itano, W.~M.} \emph{et~al.}
\newblock \bibinfo{title}{Bragg diffraction from crystallized ion plasmas}.
\newblock \emph{\bibinfo{journal}{Science}} \textbf{\bibinfo{volume}{279}},
  \bibinfo{pages}{686--689}
\newblock  (\bibinfo{year}{1998}).

\bibitem{Drewsen1998Large}
\bibinfo{author}{Drewsen, M.}, \bibinfo{author}{Brodersen, C.},
  \bibinfo{author}{Hornek\ae{}r, L.}, \bibinfo{author}{Hangst, J.} \&
  \bibinfo{author}{Schifffer, J.}
\newblock \bibinfo{title}{Large ion crystals in a linear paul trap}.
\newblock \emph{\bibinfo{journal}{Phys. Rev. Lett.}}
  \textbf{\bibinfo{volume}{81}}, \bibinfo{pages}{2878--2881}
\newblock  (\bibinfo{year}{1998}).

\bibitem{Mortensen2006Observation}
\bibinfo{author}{Mortensen, A.}, \bibinfo{author}{Nielsen, E.},
  \bibinfo{author}{Matthey, T.} \& \bibinfo{author}{Drewsen, M.}
\newblock \bibinfo{title}{Observation of three-dimensional long-range order in
  small ion coulomb crystals in an rf trap}.
\newblock \emph{\bibinfo{journal}{Phys. Rev. Lett.}}
  \textbf{\bibinfo{volume}{96}}, \bibinfo{pages}{103001}
\newblock  (\bibinfo{year}{2006}).

\bibitem{Raussendorf2007Fault}
\bibinfo{author}{Raussendorf, R.} \& \bibinfo{author}{Harrington, J.}
\newblock \bibinfo{title}{Fault-tolerant quantum computation with high
  threshold in two dimensions}.
\newblock \emph{\bibinfo{journal}{Phys. Rev. Lett.}}
  \textbf{\bibinfo{volume}{98}}, \bibinfo{pages}{190504}
\newblock  (\bibinfo{year}{2007}).

\bibitem{Raussendorf2007Topological}
\bibinfo{author}{Raussendorf, R.}, \bibinfo{author}{Harrington, J.} \&
  \bibinfo{author}{Goyal, K.}
\newblock \bibinfo{title}{Topological fault-tolerance in cluster state quantum
  computation}.
\newblock \emph{\bibinfo{journal}{New J. Phys.}} \textbf{\bibinfo{volume}{9}},
  \bibinfo{pages}{199}
\newblock  (\bibinfo{year}{2007}).

\bibitem{Fowler2009High}
\bibinfo{author}{Fowler, A.~G.}, \bibinfo{author}{Stephens, A.~M.} \&
  \bibinfo{author}{Groszkowski, P.}
\newblock \bibinfo{title}{High-threshold universal quantum computation on the
  surface code}.
\newblock \emph{\bibinfo{journal}{Phys. Rev. A}} \textbf{\bibinfo{volume}{80}},
  \bibinfo{pages}{052312}
\newblock  (\bibinfo{year}{2009}).

\bibitem{DiVincenzo2009Fault}
\bibinfo{author}{DiVincenzo, D.~P.}
\newblock \bibinfo{title}{Fault-tolerant architectures for superconducting
  qubits}.
\newblock \emph{\bibinfo{journal}{Physica Scripta}}
  \textbf{\bibinfo{volume}{2009}}, \bibinfo{pages}{014020}
\newblock  (\bibinfo{year}{2009}).

\bibitem{Cirac2000Scalable}
\bibinfo{author}{Cirac, J.~I.} \& \bibinfo{author}{Zoller, P.}
\newblock \bibinfo{title}{A scalable quantum computer with ions in an array of
  microtraps}.
\newblock \emph{\bibinfo{journal}{Nature}} \textbf{\bibinfo{volume}{404}},
  \bibinfo{pages}{579--581}
\newblock  (\bibinfo{year}{2000}).

\bibitem{Porras2006Quantum}
\bibinfo{author}{Porras, D.} \& \bibinfo{author}{Cirac, J.~I.}
\newblock \bibinfo{title}{Quantum manipulation of trapped ions in two
  dimensional coulomb crystals}.
\newblock \emph{\bibinfo{journal}{Phys. Rev. Lett.}}
  \textbf{\bibinfo{volume}{96}}, \bibinfo{pages}{250501}
\newblock  (\bibinfo{year}{2006}).

\bibitem{Zou2010Implementation}
\bibinfo{author}{Zou, P.}, \bibinfo{author}{Xu, J.}, \bibinfo{author}{Song, W.}
  \& \bibinfo{author}{Zhu, S.-L.}
\newblock \bibinfo{title}{{Implementation of local and high-fidelity quantum
  conditional phase gates in a scalable two-dimensional ion trap.}}
\newblock \emph{\bibinfo{journal}{Phys. Lett. A}}
  \textbf{\bibinfo{volume}{374}}, \bibinfo{pages}{1425--1430}
\newblock  (\bibinfo{year}{2010}).

\bibitem{Mitchell1998Direct}
\bibinfo{author}{Mitchell, T.~B.} \emph{et~al.}
\newblock \bibinfo{title}{Direct observations of structural phase transitions
  in planar crystallized ion plasmas}.
\newblock \emph{\bibinfo{journal}{Science}} \textbf{\bibinfo{volume}{282}},
  \bibinfo{pages}{1290--1293}
\newblock  (\bibinfo{year}{1998}).

\bibitem{Kielpinski2002Architecture}
\bibinfo{author}{Kielpinski, D.}, \bibinfo{author}{Monroe, C.} \&
  \bibinfo{author}{Wineland, D.~J.}
\newblock \bibinfo{title}{Architecture for a large-scale ion-trap quantum
  computer}.
\newblock \emph{\bibinfo{journal}{Nature}} \textbf{\bibinfo{volume}{417}},
  \bibinfo{pages}{709--711}
\newblock  (\bibinfo{year}{2002}).

\bibitem{Monroe2013Scaling}
\bibinfo{author}{Monroe, C.} \& \bibinfo{author}{Kim, J.}
\newblock \bibinfo{title}{Scaling the ion trap quantum processor}.
\newblock \emph{\bibinfo{journal}{Science}} \textbf{\bibinfo{volume}{339}},
  \bibinfo{pages}{1164--1169}
\newblock  (\bibinfo{year}{2013}).

\bibitem{berkeland1998minimization}
\bibinfo{author}{Berkeland, D.}, \bibinfo{author}{Miller, J.},
  \bibinfo{author}{Bergquist, J.}, \bibinfo{author}{Itano, W.} \&
  \bibinfo{author}{Wineland, D.}
\newblock \bibinfo{title}{Minimization of ion micromotion in a paul trap}.
\newblock \emph{\bibinfo{journal}{J. Appl. Phys.}}
  \textbf{\bibinfo{volume}{83}}, \bibinfo{pages}{5025--5033}
\newblock  (\bibinfo{year}{1998}).

\bibitem{Leibfried2003Quantum}
\bibinfo{author}{Leibfried, D.}, \bibinfo{author}{Blatt, R.},
  \bibinfo{author}{Monroe, C.} \& \bibinfo{author}{Wineland, D.}
\newblock \bibinfo{title}{Quantum dynamics of single trapped ions}.
\newblock \emph{\bibinfo{journal}{Rev. Mod. Phys.}}
  \textbf{\bibinfo{volume}{75}}, \bibinfo{pages}{281--324}
\newblock  (\bibinfo{year}{2003}).

\bibitem{Shen2014High}
\bibinfo{author}{Shen, C.} \& \bibinfo{author}{Duan, L.-M.}
\newblock \bibinfo{title}{High-fidelity quantum gates for trapped ions under
  micromotion}.
\newblock \emph{\bibinfo{journal}{Phys. Rev. A}} \textbf{\bibinfo{volume}{90}},
  \bibinfo{pages}{022332}
\newblock  (\bibinfo{year}{2014}).

\bibitem{Landa2012Modes}
\bibinfo{author}{Landa, H.}, \bibinfo{author}{Drewsen, M.},
  \bibinfo{author}{Reznik, B.} \& \bibinfo{author}{Retzker, A.}
\newblock \bibinfo{title}{Modes of oscillation in radiofrequency paul traps}.
\newblock \emph{\bibinfo{journal}{New J. Phys.}} \textbf{\bibinfo{volume}{14}},
  \bibinfo{pages}{093023}
\newblock  (\bibinfo{year}{2012}).

\bibitem{Kaufmann2012Precise}
\bibinfo{author}{Kaufmann, H.} \emph{et~al.}
\newblock \bibinfo{title}{Precise experimental investigation of eigenmodes in a
  planar ion crystal}.
\newblock \emph{\bibinfo{journal}{Phys. Rev. Lett.}}
  \textbf{\bibinfo{volume}{109}}, \bibinfo{pages}{263003}
\newblock  (\bibinfo{year}{2012}).

\bibitem{Landa2014Entanglement}
\bibinfo{author}{Landa, H.}, \bibinfo{author}{Retzker, A.},
  \bibinfo{author}{Schaetz, T.} \& \bibinfo{author}{Reznik, B.}
\newblock \bibinfo{title}{Entanglement generation using discrete solitons in
  coulomb crystals}.
\newblock \emph{\bibinfo{journal}{Phys. Rev. Lett.}}
  \textbf{\bibinfo{volume}{113}}, \bibinfo{pages}{053001}
\newblock  (\bibinfo{year}{2014}).

\bibitem{Yoshimura2014Creation}
\bibinfo{author}{{Yoshimura}, B.}, \bibinfo{author}{{Stork}, M.},
  \bibinfo{author}{{Dadic}, D.}, \bibinfo{author}{{Campbell}, W.~C.} \&
  \bibinfo{author}{{Freericks}, J.~K.}
\newblock \bibinfo{title}{{Creation of two-dimensional coulomb crystals of ions
  in oblate Paul traps for quantum simulations}}.
\newblock \emph{\bibinfo{journal}{ArXiv e-prints}}
\newblock  (\bibinfo{year}{2014}).
\newblock \eprint{1406.5545}.

\bibitem{Zhu2006Arbitrary}
\bibinfo{author}{Zhu, S.-L.}, \bibinfo{author}{Monroe, C.} \&
  \bibinfo{author}{Duan, L.-M.}
\newblock \bibinfo{title}{Arbitrary-speed quantum gates within large ion
  crystals through minimum control of laser beams}.
\newblock \emph{\bibinfo{journal}{Europhys. Lett.}}
  \textbf{\bibinfo{volume}{73}}, \bibinfo{pages}{485}
\newblock  (\bibinfo{year}{2006}).

\bibitem{Zhu2006Trapped}
\bibinfo{author}{Zhu, S.-L.}, \bibinfo{author}{Monroe, C.} \&
  \bibinfo{author}{Duan, L.-M.}
\newblock \bibinfo{title}{Trapped ion quantum computation with transverse
  phonon modes}.
\newblock \emph{\bibinfo{journal}{Phys. Rev. Lett.}}
  \textbf{\bibinfo{volume}{97}}, \bibinfo{pages}{050505}
\newblock  (\bibinfo{year}{2006}).

\bibitem{Choi2014Optimal}
\bibinfo{author}{Choi, T.} \emph{et~al.}
\newblock \bibinfo{title}{Optimal quantum control of multimode couplings
  between trapped ion qubits for scalable entanglement}.
\newblock \emph{\bibinfo{journal}{Phys. Rev. Lett.}}
  \textbf{\bibinfo{volume}{112}}, \bibinfo{pages}{190502}
\newblock  (\bibinfo{year}{2014}).

\bibitem{mclachlan1951theory}
\bibinfo{author}{McLachlan, N.~W.}
\newblock \emph{\bibinfo{title}{Theory and application of Mathieu functions}}
\newblock  (\bibinfo{publisher}{Clarendon Press}, \bibinfo{year}{1951}).

\bibitem{king1999quantum}
\bibinfo{author}{King, B.~E.}
\newblock \emph{\bibinfo{title}{Quantum state engineering and information
  processing with trapped ions}}.
\newblock Ph.D. thesis, \bibinfo{school}{University of Colorado}
\newblock  (\bibinfo{year}{1999}).

\bibitem{james1998quantum}
\bibinfo{author}{James, D.}
\newblock \bibinfo{title}{Quantum dynamics of cold trapped ions with
  application to quantum computation}.
\newblock \emph{\bibinfo{journal}{Appl. Phys. B}}
  \textbf{\bibinfo{volume}{66}}, \bibinfo{pages}{181--190}
\newblock  (\bibinfo{year}{1998}).

\bibitem{Zhang2007Molecular}
\bibinfo{author}{Zhang, C.~B.}, \bibinfo{author}{Offenberg, D.},
  \bibinfo{author}{Roth, B.}, \bibinfo{author}{Wilson, M.~A.} \&
  \bibinfo{author}{Schiller, S.}
\newblock \bibinfo{title}{Molecular-dynamics simulations of cold single-species
  and multispecies ion ensembles in a linear paul trap}.
\newblock \emph{\bibinfo{journal}{Phys. Rev. A}} \textbf{\bibinfo{volume}{76}},
  \bibinfo{pages}{012719}
\newblock  (\bibinfo{year}{2007}).

\bibitem{Schiffer2000Temperature}
\bibinfo{author}{Schiffer, J.~P.}, \bibinfo{author}{Drewsen, M.},
  \bibinfo{author}{Hangst, J.~S.} \& \bibinfo{author}{Hornek{\ae}r, L.}
\newblock \bibinfo{title}{Temperature, ordering, and equilibrium with
  time-dependent confining forces}.
\newblock \emph{\bibinfo{journal}{Proc. Natl. Acad. Sci.}}
  \textbf{\bibinfo{volume}{97}}, \bibinfo{pages}{10697--10700}
\newblock  (\bibinfo{year}{2000}).

\bibitem{Leibfried2003Experimental}
\bibinfo{author}{Leibfried, D.} \emph{et~al.}
\newblock \bibinfo{title}{Experimental demonstration of a robust, high-fidelity
  geometric two ion-qubit phase gate}.
\newblock \emph{\bibinfo{journal}{Nature}} \textbf{\bibinfo{volume}{422}},
  \bibinfo{pages}{412--415}
\newblock  (\bibinfo{year}{2003}).

\bibitem{Kim2009Entanglement}
\bibinfo{author}{Kim, K.} \emph{et~al.}
\newblock \bibinfo{title}{Entanglement and tunable spin-spin couplings between
  trapped ions using multiple transverse modes}.
\newblock \emph{\bibinfo{journal}{Phys. Rev. Lett.}}
  \textbf{\bibinfo{volume}{103}}, \bibinfo{pages}{120502}
\newblock  (\bibinfo{year}{2009}).

\bibitem{Lin2009Large}
\bibinfo{author}{Lin, G.-D.} \emph{et~al.}
\newblock \bibinfo{title}{Large-scale quantum computation in an anharmonic
  linear ion trap}.
\newblock \emph{\bibinfo{journal}{Europhys. Lett.}}
  \textbf{\bibinfo{volume}{86}}, \bibinfo{pages}{60004}
\newblock  (\bibinfo{year}{2009}).

\bibitem{Ryjkov2005Simulations}
\bibinfo{author}{Ryjkov, V.~L.}, \bibinfo{author}{Zhao, X.} \&
  \bibinfo{author}{Schuessler, H.~A.}
\newblock \bibinfo{title}{Simulations of the rf heating rates in a linear
  quadrupole ion trap}.
\newblock \emph{\bibinfo{journal}{Phys. Rev. A}} \textbf{\bibinfo{volume}{71}},
  \bibinfo{pages}{033414}
\newblock  (\bibinfo{year}{2005}).

\bibitem{Buluta2008Investigation}
\bibinfo{author}{Buluta, I.}, \bibinfo{author}{Kitaoka, M.},
  \bibinfo{author}{Georgescu, S.} \& \bibinfo{author}{Hasegawa, S.}
\newblock \bibinfo{title}{Investigation of planar coulomb crystals for quantum
  simulation and computation}.
\newblock \emph{\bibinfo{journal}{Phys. Rev. A}} \textbf{\bibinfo{volume}{77}},
  \bibinfo{pages}{062320}
\newblock  (\bibinfo{year}{2008}).

\end{thebibliography}

\mbox{}\newline
{\large \textbf{Acknowledgments}}\\
We would like to thank T. Choi and Z.-X. Gong for useful discussions. This work was supported by the NBRPC (973 Program) No.\ 2011CBA00300 (No.\ 2011CBA00302), the IARPA MUSIQC program, the ARO, and the AFOSR MURI program. \\
\newline
{\large \textbf{Author contributions}}\\
C.S. and L.-M.D. conceived the idea. S.-T.W. and C.S. carried out the calculations. S.-T.W. and L.-M.D. wrote the manuscript. All authors contributed to the discussion of the project and revision of the manuscript.  \\
\newline
{\large \textbf{Additional information}}\\
\textbf{Supplementary information} is available.\\
\textbf{Competing financial interests:} The authors declare no competing financial interests.

\clearpage

\onecolumngrid

\section{Supplementary Information: Quantum Computation under Micromotion in a Planar Ion Crystal}

\begin{quote}
In this supplementary information, we provide more details on the iterative method to find dynamic ion positions, and also consider the effect of in-plane micromotion to the transverse normal modes. We also include a more detailed derivation for the Hamiltonian and time-evolution operator for a two-ion entangling gate. 
\end{quote}

\section{Iterative method to find dynamic ion positions}\label{App:Iterative}

\begin{figure}[t]
\hspace{-.1cm}\includegraphics[trim=11.1cm .1cm 13.1cm 1.5cm, clip,width=0.4\textwidth]{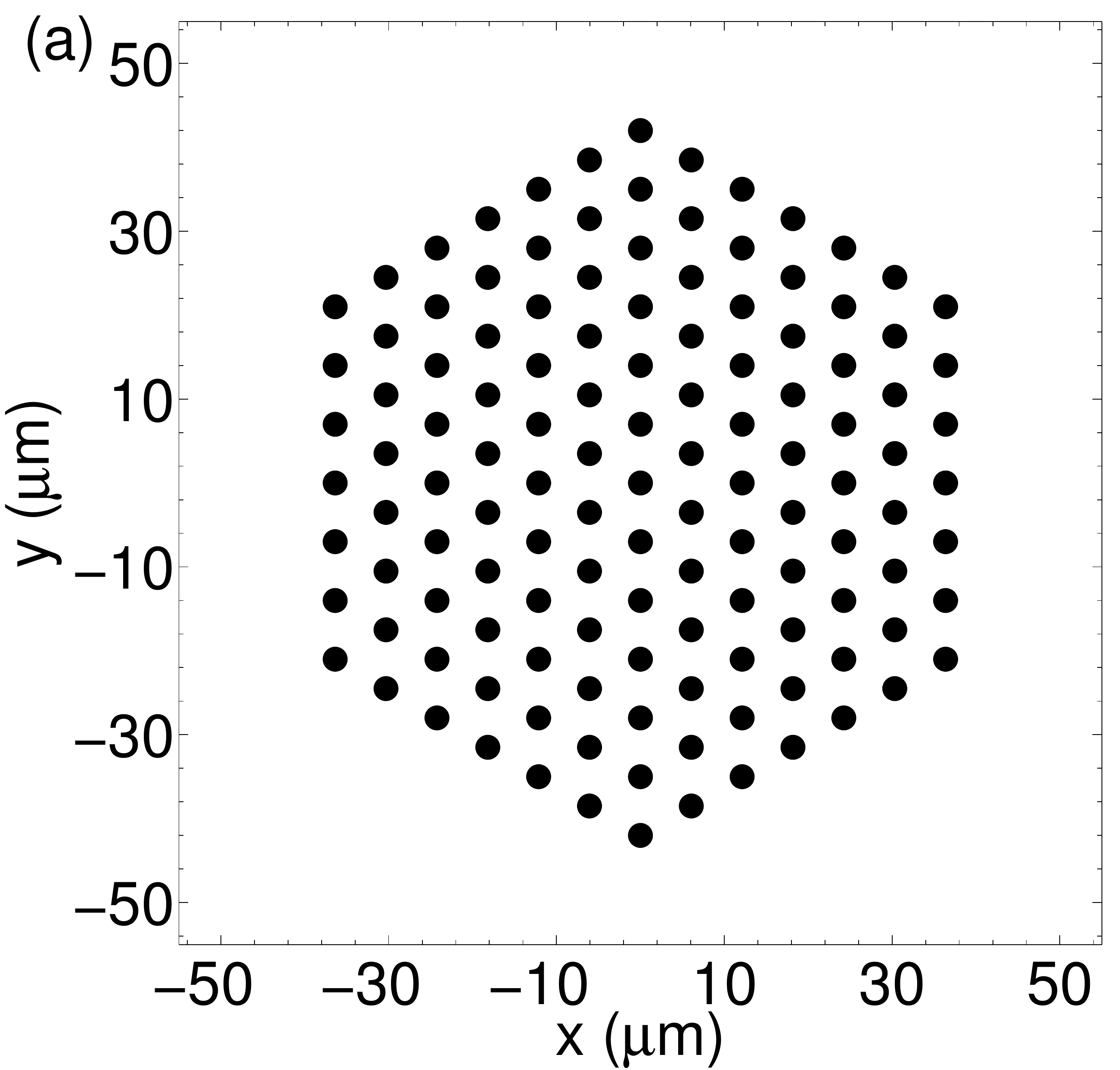}
\hspace{-.12cm}\includegraphics[trim=11.1cm .1cm 13.1cm 1.5cm, clip,width=0.4\textwidth]{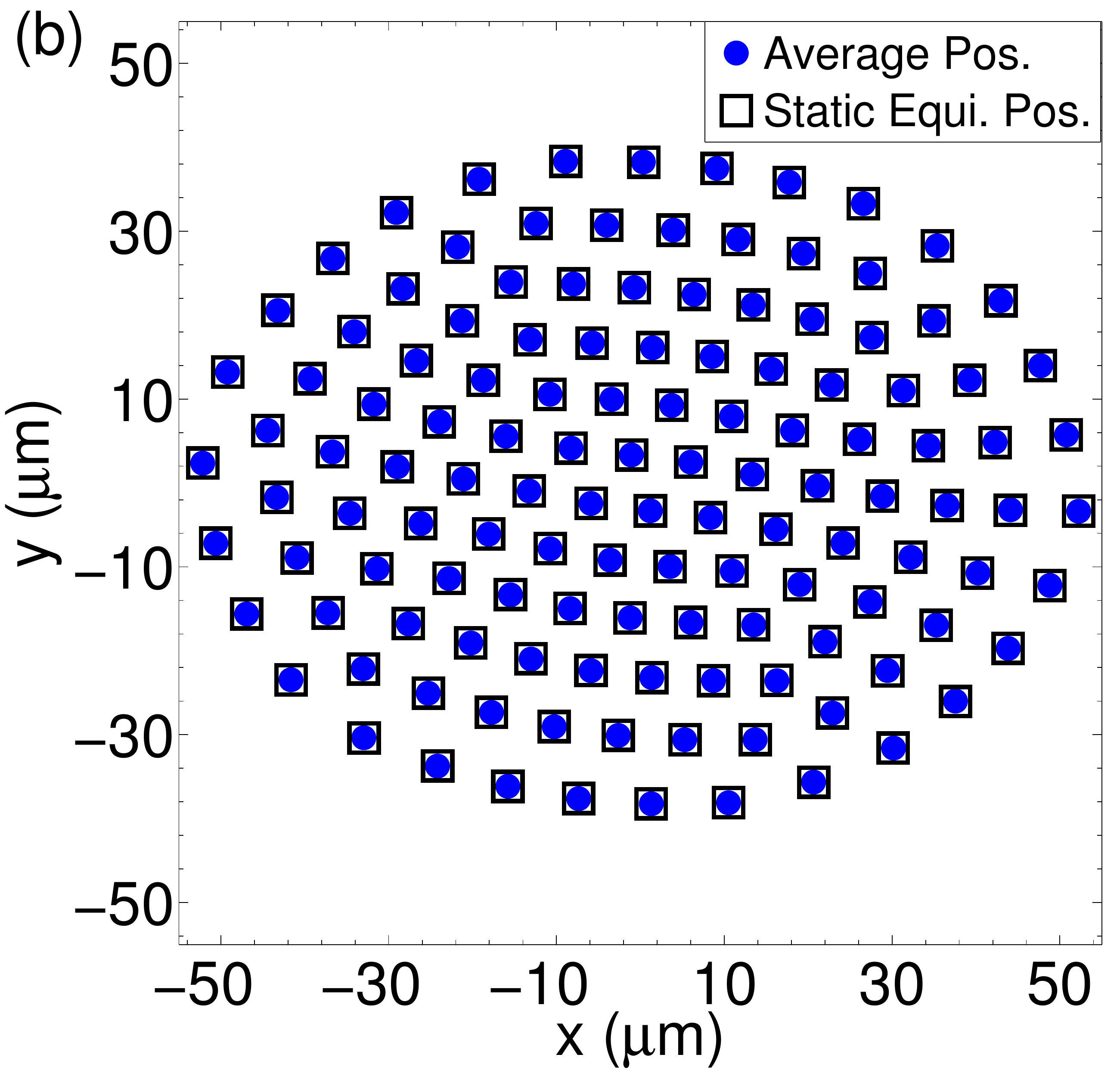}
\caption{(a) Initial configuration for ion crystal. 127 ions forming equilateral triangles with ion distance $7 \, \mu$m are arranged in a 2D hexagonal structure. (b) Stable ion configuration under the trap and Coulomb potential. Static equilibrium positions under the pseudopotential approximation are marked by (black) squares. Average ion positions found self-consistently by solving the Mathieu equations are marked by (blue) dots. The difference between two sets of equilibrium positions is around $0.03 \, \mu m$ on average, which is hardly visible in the figure.}
\label{FigSup:IonPosition}
\end{figure}

As discussed in the main text, the equations of motion in each direction can be written in the standard form of Mathieu equations (neglecting Coulomb potential):
\begin{equation}
\frac{d^{2}r_{\nu}}{d\xi^{2}}+\left[a_{\nu}-2q_{\nu}\cos(2\xi)\right]r_{\nu}=0,
\end{equation}
where $\nu \in \{x,y,z \}$, $\xi=\Omega_{T} t/2$, and dimensionless parameters $a_{\nu}$ and $q_{\nu}$ are defined in the main text.
The characteristic exponents $\beta_{\nu}$ can be computed from $a_{\nu}$ and $q_{\nu}$ iteratively \cite{mclachlan1951theory}. A pseudopotential can then be obtained with secular frequencies $\omega_{\nu} = \beta_{\nu} \Omega_{T}/2$ and
\begin{equation}
e\left(\Phi_{\text{DC}}+\Phi_{\text{AC}}\right)\approx\frac{1}{2}m\omega_{x}^{2}x^{2}+\frac{1}{2}m\omega_{y}^{2}y^{2}+\frac{1}{2}m\omega_{z}^{2}z^{2}.
\end{equation}
Assuming tight trapping along the $z$ direction, i.e.\ $\omega_{z}/\omega_{x,y}>10$, a planar crystal is formed in the $x$-$y$ plane. Adding the Coulomb potential $V_{C}$, one acquires a time-independent potential in the plane:
\begin{align}
V_{\text{pseudo}}(x,\, y) =\sum_{i}\left(\frac{1}{2}m\omega_{x}^{2}x_{i}^{2}+\frac{1}{2}m\omega_{y}^{2}y_{i}^{2}\right)+ 
\sum_{i<j}  \frac{e^{2}}{4\pi\epsilon_{0}\sqrt{(x_{i}-x_{j})^{2}+(y_{i}-y_{j})^{2}}}.
\end{align}
$i=1,2,\cdots,N$, where $N$ is the number of ions. Numerically, we start with $N=127$ ions forming equilateral triangles in a 2D hexagonal structure [Fig.\ \ref{FigSup:IonPosition}(a)], and find the static equilibrium positions $\vec{r}\,^{(0)} = (x_{1}^{(0)},y_{1}^{(0)}, \cdots, x_{N}^{(0)},y_{N}^{(0)})$ under this pseudopotential approximation by solving the classical equations of motion with a frictional force $\left( -\eta(\dot{x}+\dot{y}) \right)$, simulating the cooling process in experiment.
This set of static equilibrium positions [marked by squares in Fig.\ \ref{FigSup:IonPosition}(b)] is the starting point to derive the oscillatory behavior of each ion under micromotion.

In a planar crystal, the ions oscillate slightly around their average positions, so it is appropriate to expand the Coulomb potential around the equilibrium positions $\vec{r}\,^{(0)}$. To the second order, the Coulomb potential can be written in a quadratic form:
\begin{equation}
V_{C} \approx \dfrac{1}{2} \vec{r}\,^{T} M_{C} \vec{r} + \vec{g} \,^{T} \vec{r} + \text{constant term}, \label{EqSup:Expansion}
\end{equation}
where $\vec{r}=(x_{1},y_{1}, \cdots, x_{N}, y_{N})$, $M_{C}$ is a $2N \times 2N$ matrix, and $\vec{g}$ is a $2N$-vector. The trapping potential can also be written in this coordinate basis:
\begin{equation}
e\left(\Phi_{\text{DC}}+\Phi_{\text{AC}}\right) = \dfrac{1}{2} \vec{r}\,^{T} M_{DC} \vec{r} + \frac{V_{0}}{d_{0}^{2}} \cos(\Omega_{T} t)  \vec{r}\,^{T} I_{2N} \,\vec{r},
\end{equation}
where $I_{2N}$ is the $2N \times 2N$ identity matrix, and $M_{DC}$ is a diagonal matrix with $2(1+\gamma) eU_{0}/d_{0}^{2}$ in the odd rows (x coordinates), and $2(1-\gamma) eU_{0}/d_{0}^{2}$ in the even rows (y coordinates). Therefore, the total potential energy is
\begin{equation}
V =  \dfrac{1}{2} \vec{r}\,^{T} (M_{DC}+M_{C}) \vec{r} + \frac{V_{0}}{d_{0}^{2}}\cos(\Omega_{T} t)  \vec{r}\,^{T} I_{2N} \,\vec{r} + \vec{g} \,^{T} \vec{r}.
\end{equation}
Note that the time-dependent part of the potential is isotropic in the coordinates, so it does not couple each Mathieu equations. We can find an orthogonal matrix $Q$ that diagonalizes the first term, i.e.\ $Q(M_{DC}+M_{C}) Q^{T}= \Lambda$. Using the normal coordinates $ \vec{s} = Q  \vec{r}$, the equations of motion form decoupled Mathieu equations:
\begin{equation}
\frac{d^{2}s_{i}}{d\xi^{2}}+ (a_{i}-2q_{i} \cos(2\xi)) s_{i}= f_{i},
\end{equation}
where $a_{i}= 4\Lambda_{ii}/m\Omega_{T}^{2}$, $q_{i} =q= -4 eV_{0}/md_{0}^{2}\Omega_{T}^{2}$, and $f_{i}= -\frac{4}{m\Omega_{T}^{2}} \left(Q\vec{g} \right)_{i}$. The inhomogeneous Mathieu equations can be solved by substituting a special solution in the form of $s_{i} = f_{i} \sum_{n=0}^{\infty} c_{i}^{(n)} \cos(2n\xi)$, and the series coefficients $c_{i}^{(n)}$ can be computed numerically \cite{Shen2014High}. After that, the ion coordinates can be transformed back to the Cartesian coordinates $\vec{r} = Q^{T}  \vec{s}$, where $\vec{r}$ can be expressed successively as
\begin{equation}
\vec{r} = \vec{r}\,^{(0)} + \vec{r}\,^{(1)} \cos(2\xi) +  \vec{r}\,^{(2)} \cos(4\xi) + \cdots.
\label{EqSup:Positions}
\end{equation}
$ \vec{r}\,^{(0)}$ now becomes the new average (equilibrium) positions, and can be substituted back to the expansion in equation \eqref{EqSup:Expansion}. The ion positions  $\vec{r}$ can be attained self-consistently in this manner. A dynamical expansion of the Coulomb potential around $\vec{r}\,^{(0)} + \vec{r}\,^{(1)} \cos(2\xi)$ may yield a more accurate result for the normal modes in the plane \cite{Landa2012Modes}. For our purpose, the static expansion is sufficient as we only need accurate ion positions $\vec{r}$ to compute the normal modes along the $z$ direction. Numerically, we found that $\vec{r}\,^{(1)} \approx -\frac{q}{2} \vec{r}\,^{(0)}$ and $\vec{r}\,^{(2)} \approx \frac{q^{2}}{32} \vec{r}\,^{(0)}$, which are consistent with previous results \cite{Shen2014High,Landa2012Modes}. Hence, micromotion only results in breathing oscillations about the average positions of each ion. The further the ion is from the center of the trap, the larger the amplitude of micromotion becomes.

\begin{figure}[t]
\includegraphics[trim=2.8cm .1cm 4.4cm 1.5cm, clip,width=0.55\textwidth]{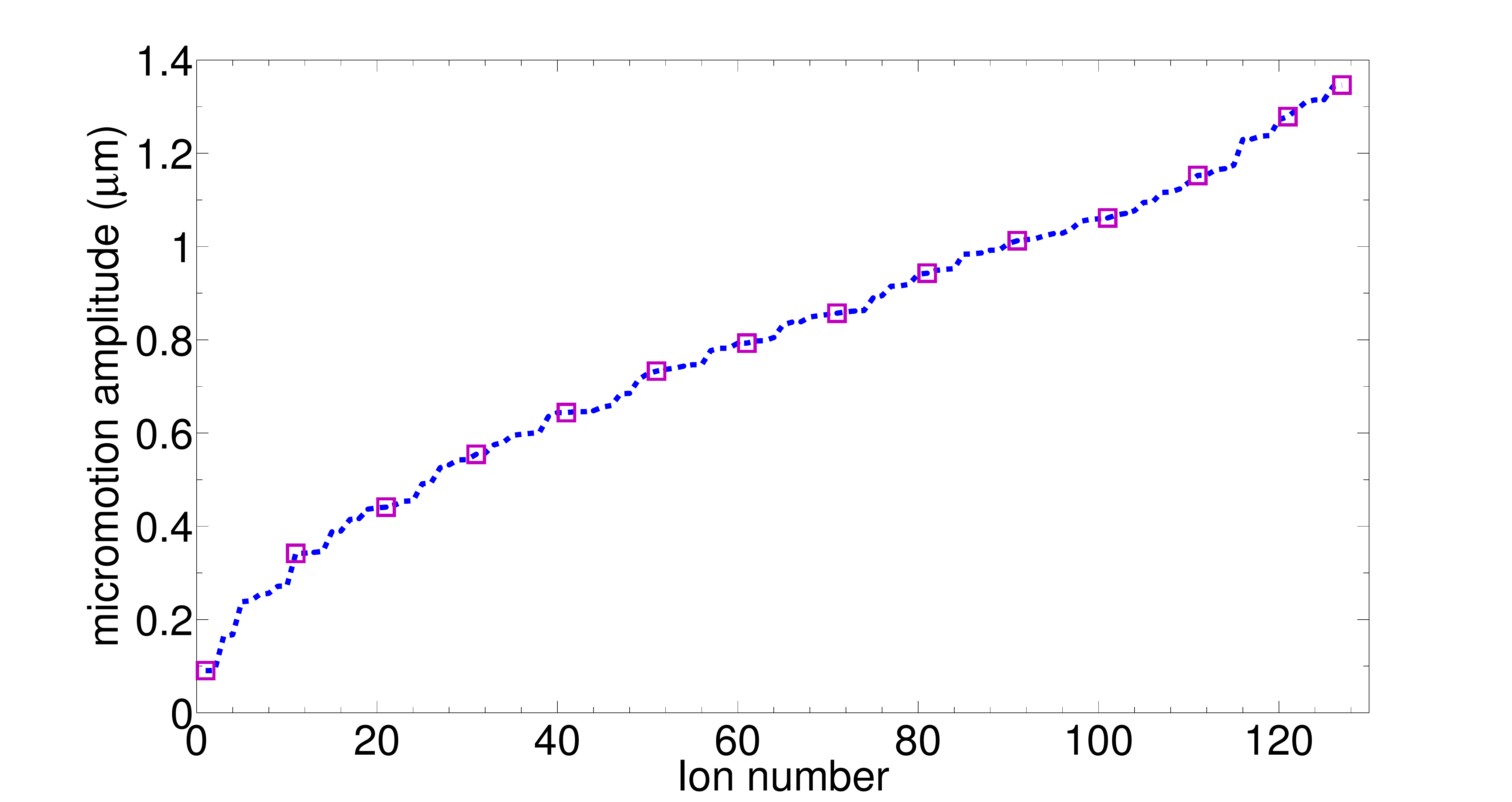}
\caption{Amplitude of micromotion for each ion (sorted in increasing order). }
\label{FigSup:MicroAmp}
\end{figure}

Fig.\ \ref{FigSup:MicroAmp} shows the amplitude of micromotion for each ion. The largest amplitude for the edge ion is around $1.35 \, \mu$m, which is well below the ion separation ($7 \sim 10 \, \mu$m), necessary for the formation of a well-defined crystal and for individual addressing.

\section{Normal modes along the transverse direction}
\label{App:Normal}

With the knowledge of the motion of ions in the $x$-$y$ plane, we could find the normal modes and quantize the motion along the transverse ($z$) direction.  As ions are confined in the plane, micromotion along the transverse direction is negligible. A harmonic pseudopotential is thus valid for the $z$ direction. Expanding the Coulomb potential to second order again, we have
\begin{align}
V_{z} = \dfrac{1}{2}m \omega_{z}^{2} \sum_{i} z_{i}^{2} +
\dfrac{e^{2}}{4\pi \epsilon_{0}} \left[ \sum_{i\neq j} \left(\dfrac{1}{r_{ij}^{3}} \right) z_{i} z_{j} - \sum_{i\neq j} \left(\dfrac{1}{r_{ij}^{3}} \right) z_{i}^{2} \right],
\end{align}
where $r_{ij}= \sqrt{(x_{i}- x_{j})^{2}+ (y_{i}-y_{j})^{2} }$. $x_{i}(t)$ and $y_{i}(t)$ are time-dependent though, due to the in-plane micromotion. From here, we can see explicitly that the transverse modes are decoupled from the planar modes. Expanding the term $1/r_{ij}^{3}(t)$ in series, one has
\begin{equation}
\dfrac{1}{r_{ij}^{3}} \approx \Big< \dfrac{1}{r_{ij}^{3}} \Big> + M_{ij} \cos (\Omega_{T} t) + \cdots
\end{equation}
The matrix element $M_{ij}$ is in the order of $O(q)$ and can be obtained numerically from $\left<\cos(\Omega_{T} t) /r_{ij}^{3} \right>$. To have an intuitive understanding of the effect of micromotion on transverse modes, we take positions $\vec{r}$ in the form of Eq.\  \eqref{EqSup:Positions}, obtaining
\begin{align}
\dfrac{1}{r_{ij}^{3}} \approx  \left( \dfrac{1}{r_{ij}^{(0)}} \right)^{3} \left(1 -\dfrac{q}{2} \cos(\Omega_{T} t) + \dfrac{q^{2}}{32} \cos(2\Omega_{T} t) \right)^{-3} + O(q^{3}),
\end{align}
where $r_{ij}^{(0)}$ is the zeroth order approximation using the average positions $ \vec{r}\,^{(0)}$ without considering micromotion. Thus, $\left<1 /r_{ij}^{3} \right> \approx \left( 1/r_{ij}^{(0)} \right)^{3}(1-3q^{2}/4) + O(q^{3})$, where we used the fact that $\left< \cos(\Omega_{T} t)\right> =0$ and $\left< \cos^{2}(\Omega_{T} t)\right> =1/2$. From the time-independent term $\left<1 /r_{ij}^{3} \right>$, we diagonalize $V_{z}$ and find the normal modes as well as the eigenenergies in the transverse direction.  Subsequently, we quantize the total Hamiltonian (with kinetic energy) and write $H = \sum_{k} \hbar \omega_{k} a_{k}^{\dagger} a_{k}$, where $a_{k}$ is the annihilation operator for the quantized phonon mode, and $\omega_{k}$ is the corresponding eigenfrequency. In the interaction picture, $a_{k} \to a_{k} e^{-i\omega_{k} t}$. The time-dependent term containing $\cos(\Omega_{T} t)$ can then be treated as a perturbation; under the rotating wave approximation, since $\Omega_{T} \gg \omega_{k}$, the term affects the normal modes to the order of $O\left(q\omega_{k}^{2}/ \Omega_{T}^{2}  \right) \sim O(qq_{z}^{2})$, which can be safely neglected. Since the first term in $V_{z}$ is diagonal in $z_{i}$ and the second term is reduced by a factor $(1-3q^{2}/4)$ by micromotion, the normal mode structure remains unchanged, and the mode frequencies are reduced slightly.

\section{Two-ion Entangling Gate}
\label{App:Entangle}

The spin-dependent force on an ion is due to the AC Stark shift on each spin state. A different shift on the two internal spin states of an ion results in a Hamiltonian
\begin{equation}
H =\hbar \dfrac{|\Omega_{\text{eg}}|^{2}}{4\delta} \sigma^{z},
\end{equation}
where $\Omega_{\text{eg}}$ is the Rabi frequency of the laser beam and $\delta$ is the detuning from the excited state. By shining two laser beams at an angle with wave vectors $\mathbf{k}_{1}$, $\mathbf{k}_{2}$ and frequencies $\omega_{1}$, $\omega_{2}$, we have 
\begin{equation}
\Omega_{\text{eg}}= \Omega_{0} \left( e^{i (\mathbf{k}_{1} \cdot \mathbf{r} + \omega_{1} t+\phi) } + e^{i (\mathbf{k}_{2} \cdot \mathbf{r} + \omega_{2} t )} \right),
\end{equation}
where $\phi$ is the phase difference between two beams. So we have 
\begin{equation}
H = \hbar \Omega \left(1+ \cos(\Delta k \cdot z + \mu t + \phi)  \right) \sigma^{z},
\end{equation}
where $\Omega = \Omega_{0}^{2}/2\delta$ is the effective two-photon Rabi frequency, $\Delta k \hat{z}= \mathbf{k}_{1} - \mathbf{k}_{2}$ is aligned along the $z$ direction, and $\mu = \omega_{1} - \omega_{2}$. As we are mostly interested in the two-qubit entangling gate, which is the building block for universal quantum gates, we consider laser beams shining on two ions, and ignore the first term $\hbar \Omega \sigma^{z}$ in the Hamiltonian that only induces single bit operations. We therefore have
\begin{equation}
H = \sum_{j=1}^{2} \hbar \Omega_{j} \cos (\Delta k \cdot z_{j} + \mu t + \phi_{j}) \sigma_{j}^{z},
\end{equation}
The ion position $z_{j} =z_{j0} + \delta z_{j}$, where $z_{j0}$ is the equilibrium position and $\delta z_{j}$ is the small displacement. We dump the term $\Delta k \cdot z_{j0}$ to the phase $\phi_{j}$, and expand the cosine term in the Lamb-Dicke limit $\Delta k \cdot \delta z_{j} \ll 1$,
\begin{align}
H &= \sum_{j=1}^{2} \hbar \Omega_{j} \cos (\Delta k \cdot \delta z_{j} + \mu t + \phi_{j}) \sigma_{j}^{z} \\
&\approx - \sum_{j=1}^{2} \hbar \Omega_{j} \sin(\Delta k \cdot \delta z_{j}) \sin ( \mu t + \phi_{j}) \sigma_{j}^{z} \label{Eq:expansion} \\
& \approx - \sum_{j,k} \hbar \Omega_{j}  \sin ( \mu t + \phi_{j}) \Delta k \left[\sqrt{\dfrac{\hbar}{2m\omega_{k}}} b_{j}^{k} a_{k}^{\dag} + \text{H.c.}  \right] \sigma_{j}^{z} \notag \\
&= - \sum_{j=1}^{2} \sum_{k} \chi_{j} (t) g_{j}^{k} (a^{\dag}_{k}+ a_{k}) \sigma_{j}^{z}
\end{align}
In step \eqref{Eq:expansion}, we drop the cosine-cosine term $\hbar \Omega_{j} \cos(\Delta k \cdot \delta z_{j}) \cos ( \mu t + \phi_{j}) \sigma_{j}^{z}  \approx \hbar \Omega_{j} \cos ( \mu t + \phi_{j}) \sigma_{j}^{z} $ since $\Delta k \cdot \delta z_{j} \ll 1$ and it thus does not couple the phonon modes to the spin (in the first-order approximation), resulting in a single-qubit operation. Various terms are defined as 
\begin{equation}
\delta z_{j} = \sum_{k} \sqrt{\dfrac{\hbar}{2m\omega_{k}}} b_{j}^{k} a_{k}^{\dag} + \text{H.c.} 
\end{equation}
where $b_{j}^{k}$ are the mode vector for mode $k$, $a_{k}^{\dag}$ creates the $k$-th phonon mode (harmonic oscillator mode). The matrix $b_{n}^{k}$ diagonalizes the approximate harmonic potential of the system.
\begin{align}
\chi_{j} (t) &= \hbar \Omega_{j} \sin (\mu t + \phi_{j}) \\
g_{j}^{k} &= \eta_{k} b_{j}^{k},  \quad \text{where} \quad \eta_{k}= \Delta k \sqrt{ \dfrac{\hbar}{2m\omega_{k}} }
\end{align}
$\eta_{k}$ is the Lamb-Dicke parameter, $\eta_{k} \ll 1$ to be valid (for the expansion). For $\Delta k=8 \mu m^{-1}$, $m=171u$ for Ytterbium, and take the transverse mode $\omega_{k} = 2\pi \times 2$MHz. We will have $\eta_{k} \approx 0.03$. Going into the interaction picture and replacing $a_{k} \to a_{k} e^{-i \omega_{k} t}$, we have
\begin{equation}
H_{\text{I}}= - \sum_{j=1}^{2} \sum_{k} \chi_{j} (t) g_{j}^{k} (a^{\dag}_{k} e^{i \omega_{k} t}+ a_{k} e^{-i \omega_{k} t}) \sigma_{j}^{z}
\end{equation}
The evolution operator can be obtained from the Hamiltonian as \cite{Zhu2006Trapped, Kim2009Entanglement} 
\begin{align}
U(\tau) &= \exp \big( i \sum_{j} \phi_{j} (\tau) \sigma_{j}^{z} + i \sum_{j<n} \phi_{jn} (\tau) \sigma_{j}^{z} \sigma_{n}^{z}   \big), \\
\phi_{j} (\tau) &=-i \sum_{k} \alpha_{j}^{k} (\tau) a_{k}^{\dag}- \alpha_{j}^{k*}(\tau) a_{k} \\
 \alpha_{j}^{k}(\tau) &= \dfrac{i}{\hbar} g_{j}^{k} \int_{0}^{\tau} \chi_{j}(t) e^{i \omega_{k}t} dt, \label{Eq:Alpha} \\
\phi_{jn}(\tau) &= \dfrac{2}{\hbar^{2}} \sum_{k} g_{j}^{k}g_{n}^{k} \int_{0}^{\tau} \int_{0}^{t_{2}} \chi_{j} (t_{2}) \chi_{n}(t_{1}) \times \sin(\omega_{k}(t_{2}-t_{1})) dt_{1} dt_{2}.
\end{align}
To obtain a two-qubit entangling gate, we need $\alpha_{j}^{k} =0$ so that the spin and phonons are disentangled at the end of the gate, and $\phi_{jn} (\tau) =\pi/4$. This is the starting point to calculate the fidelity of the gate.

\end{document}